\documentclass[twoside]{ilcws08}
\usepackage[latin1]{inputenc}
\usepackage[dvips]{graphicx,epsfig,color}
\usepackage{wrapfig,rotating}
\usepackage{amssymb,amsmath,array}

\begin{document}
\title{
%%%%   Paper title goes here  %%%%%%%%%%%%%%
Event Shape Variables at NLLA+NNLO} %%
%***********************************************************************
% AUTHORS INFORMATION AREA
%***********************************************************************
\author{Gionata Luisoni
% Optional short acknowledgment: remove next line if non-needed
\thanks{On behalf of  G.\ Dissertori,
A.~Gehrmann-De Ridder, T.~Gehrmann, E.W.N.~Glover,
G.~Heinrich and H.\ Stenzel.}
% DO NOT MODIFY THE FOLLOWING '\vspace' ARGUMENT
\vspace{.3cm}\\
% Addresses and institutions (remove "1- " in case of a single institution)
Institut f\"ur
Theoretische Physik, Universit\"at Z\"urich, \\CH-8057
Z\"urich, Switzerland
%% Remove the next three lines in case of a single institution
%\vspace{.1cm}\\
%2- Second Author's Institution - Department \\
%Address of Second Author's Institution - Country\\
}
%%***********************************************************************
% END OF AUTHORS INFORMATION AREA
%***********************************************************************

\maketitle

\begin{abstract}
In this talk~\cite{url} we report work on the matching of the next-to-leading
logarithmic approximation (NLLA) onto the fixed
next-to-next-to-leading order (NNLO) calculations for event
shape distributions in electron-positron annihilation.
Furthermore we present preliminary results on the determination of the
strong coupling constant obtained using NLLA+NNLO
predictions and ALEPH data.

\end{abstract}

\section{Introduction}
The reaction of $e^{+}e^{-}$ annihilation into three jets has played historically a very prominent role for phenomenology. It permitted for example the discovery of the gluon and the measurement of its properties and allows also a precise determination of the strong coupling constant $\alpha_{s}$, since the deviation from two-jet configurations is proportional to it. Not only jet rates, but also the shape of the single events can be studied in a systematic fashion. The so-called event shape observables became very popular mainly because they are well suited both for experimental measurement and for theoretical description since many of them are infrared and collinear safe. The main idea behind event shape variables is to parameterize the energy-momentum flow of an event, such that one can smoothly describe its shape passing from pencil-like two-jet configurations, which are a limiting case in event shapes, up to multijet final states. At LEP a set of six different event shape observables were measured in great detail: thrust $T$ (which is substituted here by $\tau = 1-T$), heavy jet mass $\rho$, wide and total jet broadening $B_W$ and $B_T$, $C$-parameter and two-to-three-jet transition parameter in the Durham algorithm $\mathrm{y}_3$. The definitions of these variables, which we denote collectively as $y$ in the following, are summarized in~\cite{hasko}. The two-jet limit of each variable is $y\to 0$. Until very recently, the theoretical state-of-the-art description of event shape distributions was based on the matching of the NLLA~\cite{resumall} onto the NLO~\cite{ERT,event} calculation. Using these predictions the largest contribution to the error in the determination of the strong coupling constant came from theoretical scale uncertainties. Recently the NNLO corrections became available. Using this new results we computed the matching of the resummed NLLA onto the fixed order NNLO.

\section{Fixed order and resummed calculations}
At NNLO the integrated fixed order differential cross section
\begin{equation*}\label{Rfixed}
R\left(y,Q,\mu\right)\,\equiv\,\frac{1}{\sigma_{{\rm had}}}\int_{0}^{y}\frac{d\sigma\left(x,Q,\mu\right)}{dx}dx\,,
\end{equation*}
is given by
\begin{equation*}\label{Rfixedexp}
R\left(y,Q,\mu\right)\,=\,1+\,\bar{\alpha}_{s}\left(\mu\right)\mathcal{A}\left(y\right)\,
+\,\bar{\alpha}_{s}^{2}\left(\mu\right)\mathcal{B}\left(y,x_\mu\right)\,+\,\bar{\alpha}_{s}^{3}\left(\mu\right)\mathcal{C}\left(y,x_\mu\right)\,,
\end{equation*}
where $\bar\alpha_s = \alpha_s/(2\pi)$ and $x_\mu = \mu/Q$.

\begin{figure}[h]
\centering
  \includegraphics[width=6.0cm]{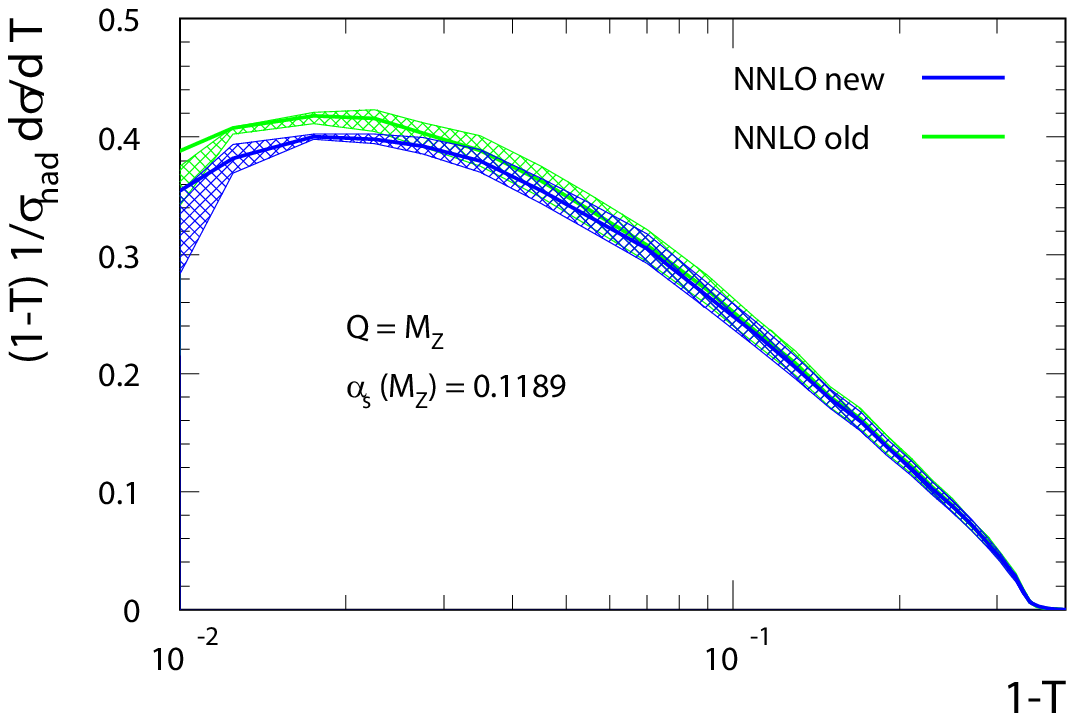}
  \qquad\qquad
  \includegraphics[width=6.0cm]{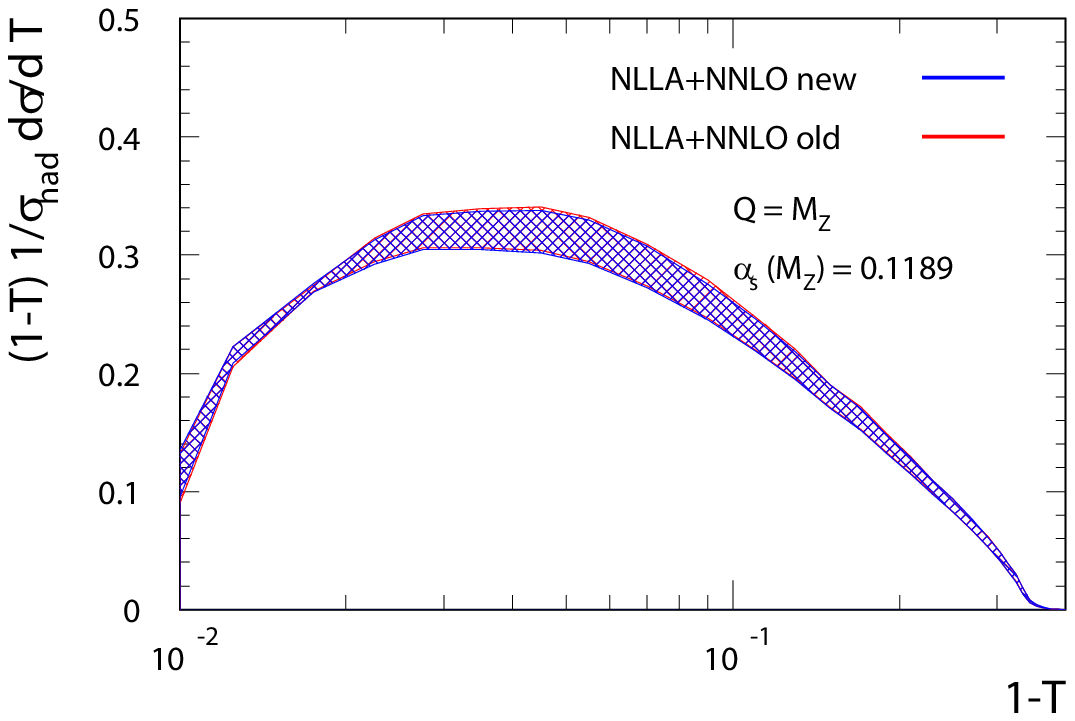}\\
  \caption{Comparison between old and corrected distributions for $\tau$. In the fixed order distribution (left) a small difference is visible in the far infrared region, in the matched distribution (right) the curves are equal since the resummation becomes dominant in the infrared region.}\label{fig:oldnewdata}
\end{figure}

\begin{wraptable}{r}{0.45\columnwidth}
\centerline{\begin{tabular}{|l|l|l|}
\hline
LO   & $\gamma^{\ast}\,\rightarrow\,q\bar{q}g$         & \textrm{tree level} \\\hline
NLO  & $\gamma^{\ast}\,\rightarrow\,q\bar{q}g$         & \textrm{one loop}   \\
     & $\gamma^{\ast}\,\rightarrow\,q\bar{q}gg$        & \textrm{tree level} \\
     & $\gamma^{\ast}\,\rightarrow\,q\bar{q}q\bar{q}$  & \textrm{tree level} \\\hline
NNLO & $\gamma^{\ast}\,\rightarrow\,q\bar{q}g$         & \textrm{two loop}   \\
     & $\gamma^{\ast}\,\rightarrow\,q\bar{q}gg$        & \textrm{one loop}   \\
     & $\gamma^{\ast}\,\rightarrow\,q\bar{q}q\bar{q}$  & \textrm{one loop}   \\
     & $\gamma^{\ast}\,\rightarrow\,q\bar{q}q\bar{q}g$ & \textrm{tree level} \\
     & $\gamma^{\ast}\,\rightarrow\,q\bar{q}ggg$       & \textrm{tree level} \\\hline
\end{tabular}}
\caption{Contributions order by order.}
\label{tab:contrib}
\end{wraptable}
\noindent
Table \ref{tab:contrib} shows the relevant contributions for the computation of the three coefficient functions $\mathcal{A}$, $\mathcal{B}$ and $\mathcal{C}$. The careful subtraction of real and virtual divergences is done using the antenna formalism and implemented in a numerical integration program. Recently an inconsistency in the treatment of large-angle soft radiation was discovered~\cite{weinzierlnew}. This was corrected (erratum to~\cite{ourirstructure}) and it results in numerically minor changes to the NNLO coefficients in the kinematical region of phenomenological interest here. The corrections turn out to be significant only in the deep two-jet region, e.g. $(1-T)<0.05$ (figure~\ref{fig:oldnewdata}).
Approaching the two-jet region the infrared logarithms in the coefficient functions become large spoiling the convergence of the series expansion. The main contribution in this case comes from the highest powers of the logarithms which have to be resummed to all orders. For suitable observables resummation leads to exponentiation. At NLLA the resummed expression is given by
\begin{equation*}\label{eq:Rresummed}
R\left(y,Q,\mu\right)\,=\ \left(1+C_{1}\bar{\alpha}_{s}
\right)\,e^{\left(L\,g_{1}\left(\alpha_{s}L\right)
+g_{2}\left(\alpha_{s}L\right)\right)}\;,
\end{equation*}
where the function $g_{1}\left(\alpha_{s}L\right)$ contains all leading-logarithms (LL), $g_{2}\left(\alpha_{s}L\right)$ all next-to-leading-logarithms (NLL) and $\mu=Q$ is used. Terms beyond NLL have been consistently omitted.
\begin{table}[h]
\centering\Large
\begin{tabular}{|l|c|c|c|c|c|c|}
       \hline
        ${\bar{\alpha}_{s}\mathcal{A}\left(y\right)}$ & $\color{blue}\bar{\alpha}_{s}L$ & $\color{red}\bar{\alpha}_{s}L^{2}$ &  &  &  &  \\\hline
        ${\bar{\alpha}_{s}^{2}\mathcal{B}\left(y,x_\mu\right)}$ & $\bar{\alpha}_{s}^{2}L$ & $\color{blue}\bar{\alpha}_{s}^{2}L^{2}$ & $\color{red}\bar{\alpha}_{s}^{2}L^{3}$ & $\color{green}\bar{\alpha}_{s}^{2}L^{4}$ &  & \\\hline
        ${\bar{\alpha}_{s}^{3}\mathcal{C}\left(y,x_\mu\right)}$ & $\bar{\alpha}_{s}^{3}L$ & $\bar{\alpha}_{s}^{3}L^{2}$ & $\color{blue}\bar{\alpha}_{s}^{3}L^{3}$ & $\color{red}\bar{\alpha}_{s}^{3}L^{4}$ & $\color{green}\bar{\alpha}_{s}^{3}L^{5}$ & $\color{green}\bar{\alpha}_{s}^{3}L^{6}$\\\hline
\end{tabular}
\caption{Powers of the logarithms present at different
orders in perturbation theory. The color highlights the
different orders in resummation: LL (red) and NLL (blue).
The terms in green are contained in the LL and NLL
contributions and exponentiate trivially with
them.}\label{tab:logs}
\end{table}
\noindent
The resummation functions $g_1(\alpha_s L)$ and $g_2(\alpha_s L)$ can be expanded as power series in $\bar{\alpha}_{s}L$:
\begin{eqnarray}\label{eq:gexpand}
L\,g_{1}\left(\alpha_{s}L\right)&=&\,G_{12}\bar{\alpha}_{s}L^{2}+G_{23}\bar{\alpha}_{s}^{2}L^{3}+G_{34}\bar{\alpha}_{s}^{3}L^{4}+\dots\;\textrm{(LL)\,,}\nonumber\\
g_{2}\left(\alpha_{s}L\right)&=&\,G_{11}\bar{\alpha}_{s}L+G_{22}\bar{\alpha}_{s}^{2}L^{2}+G_{33}\bar{\alpha}_{s}^{3}L^{3}+\dots\;\textrm{(NLL)\,.}
\end{eqnarray}
Table~\ref{tab:logs} shows the logarithmic terms present up to the third order in perturbation theory. At the fixed order level the LL are terms of the form $\alpha_{s}^{n}L^{n+1}$, the NLL those which go like $\alpha_{s}^{n}L^{n}$, and so on. Notice that this can be read off the expansion (\ref{eq:gexpand}) of the exponentiated resummation functions.

Closed analytic forms for the functions $g_1(\alpha_s L)$ and $g_2(\alpha_s L)$ are available for $\tau$ and $\rho$~\cite{resumt}, $B_W$ and $B_T$~\cite{resumbwbt,resumbwbtrecoil}, $C$~\cite{resumc} and $Y_3$~\cite{resumy3a}, and are collected in the appendix of~\cite{ourpaper}. Recently also $g_{3}\left(\alpha_{s}L\right)$ and $g_{4}\left(\alpha_{s}L\right)$ were computed for $\tau$ using effective field theory methods~\cite{scetthrust}.
\section{Matching of fixed order and resummed calculations}
To obtain a reliable description of the event shape distributions over a wide range in $y$, it is mandatory to combine fixed order and resummed predictions. The two predictions have to be matched in a way that avoids the double counting of terms present in both. A number of different matching procedures have been proposed in the literature, see for example~\cite{hasko} for a review. In the so-called $R$-matching scheme, the two expressions for $R\left(y\right)$ are matched. We computed the matching in the so-called $\ln\, R$-matching~\cite{resumall} since in this particular scheme, all matching coefficients can be extracted analytically from the resummed calculation. The $\ln\,R$-matching at NLO is described in detail in~\cite{resumall}. In the $\ln\,R$-matching scheme, the NLLA+NNLO expression is
\begin{eqnarray}\label{logRmatching}
\ln\left(R\left(y,\alpha_{s}\right)\right)&=&L\,g_{1}\left(\alpha_{s}L\right)\,+\,g_{2}\left(\alpha_{s}L\right)+\,\bar{\alpha}_{S}\left(\mathcal{A}\left(y\right)-G_{11}L-G_{12}L^{2}\right)+{}\nonumber\\
&&+\,\bar{\alpha}_{S}^{2}\left(\mathcal{B}\left(y\right)-\frac{1}{2}\mathcal{A}^{2}\left(y\right)-G_{22}L^{2}-G_{23}L^{3}\right){}\nonumber\\
&&+\,\bar{\alpha}_{S}^{3}\left(\mathcal{C}\left(y\right)-\mathcal{A}\left(y\right)\mathcal{B}\left(y\right)+\frac{1}{3}\mathcal{A}^{3}\left(y\right)-G_{33}L^{3}-G_{34}L^{4}\right)\;.
\end{eqnarray}
The matching coefficients appearing in this expression can be obtained from (\ref{eq:gexpand}) and are listed in~\cite{ourpaper}. To ensure the vanishing of the matched expression at the kinematical boundary $y_{\textrm{\tiny{max}}}$ a further shift of the logarithm is made~\cite{hasko}.

The renormalisation scale dependence of (\ref{logRmatching}) is given by making the following replacements:
\begin{eqnarray*}
\alpha_s & \to & \alpha_s(\mu)\;, \\
\mathcal{B}\left(y\right) &\to &
\mathcal{B}\left(y,\mu\right)=2\,\beta_{0}\, \ln x_\mu \,
\mathcal{A}\left(y\right)
+\mathcal{B}\left(y\right)\;,\nonumber \\
\mathcal{C}\left(y\right) & \to &
\mathcal{C}\left(y,\mu\right)=\left(2\,\beta_{0}\, \ln x_\mu
\right)^{2}\mathcal{A}\left(y\right) +2\,\ln x_\mu
\,\left[2\,\beta_{0}
\mathcal{B}\left(y\right)+2\,\beta_{1}\,\mathcal{A}\left(y\right)\right]
+\mathcal{C}\left(y\right)\;,
\label{fixedorderrenscaledependence}\\
g_2\left(\alpha_{s}L\right) &\to &
{g}_{2}\left(\alpha_{s}L,\mu^{2}\right)
=g_{2}\left(\alpha_{s}L\right)+\frac{\beta_{0}}{\pi}
\left(\alpha_{s}L\right)^{2}\, g_{1}'\left(\alpha_{s}L\right)\,\ln
x_\mu \;,
\label{g2mudep} \\
G_{22}&\to & G_{22}\left(\mu\right)=G_{22}\,+\,2\beta_{0}G_{12}\ln
x_\mu
\;,\nonumber\\
G_{33}&\to & {G}_{33}\left(\mu\right)=G_{33}\, +\,4 \beta_{0}
G_{23}\ln x_\mu\,. \label{Gijdeponrenorm}
\end{eqnarray*}
In the above, $g_1'$ denotes the derivative of $g_1$ with respect to its argument. The LO coefficient ${\cal A}$ and the LL resummation function $g_1$, as well as the matching coefficients $G_{i\,i+1}$ remain independent on $\mu$.

\section{Discussion of the matched distribution}
For the resulting plots of the matched distributions we refer to~\cite{ourpaper}. The most striking observation is that the difference between NLLA+NNLO and NNLO is largely restricted to the two-jet region, while NLLA+NLO and NLO differ in normalisation throughout the full kinematical range. This behavior may serve as a first indication for the numerical smallness of corrections beyond NNLO in the three-jet region. In the approach to the two-jet region, the NLLA+NLO and NLLA+NNLO predictions agree by construction, since the matching suppresses any fixed order terms. Although not so visible on these plots, the difference between NLLA+NNLO and NLLA+NLO is only moderate in the three-jet region. The renormalisation scale uncertainty in the three-jet region is reduced by 20-40\% between NLLA+NLO and NLLA+NNLO. This effect is due to the smaller renormalization scale dependence of the NNLO contributions. It is also important to observe that the scale dependence remains the same and is larger in the two-jet region, because the resummed calculations at NLLA take into account only the one-loop running of the coupling constant. This has important consequences in the determination of $\alpha_{s}$ and we will comment more on this in the next section.

The description of the hadron-level data improves between parton-level NLLA+NLO and parton-level NLLA+NNLO, especially in the three-jet region. The behavior in the two-jet region is described better by the resummed predictions than by the fixed order NNLO, although the agreement is far from perfect. This discrepancy can in part be attributed to missing higher order logarithmic corrections and in part to non-perturbative corrections, which become large in the approach to the two-jet limit.

The right plot in figure~\ref{fig:oldnewdata} shows that the inconsistency in the treatment of the large-angle soft radiation does not affect the matched prediction since the infrared region is dominated by the resummation.

\section{Determination of the strong coupling constant}

After the extraction of $\alpha_{s}$ using only the NNLO distributions and the experimental data of ALEPH~\cite{ouralphas}, a new extraction of $\alpha_s$ using the new matched results was performed using JADE data~\cite{alphasfitmunich}. The improvement in the error coming from the inclusion of resummed calculation is not as drammatic as passing from NLO to NLLA+NLO calculations.
As already anticipated, this is due to the fact that the NNLO coefficients compensate the two-loop renormalization scale variation, whereas the NLLA part only compensates the one-loop variation. A more natural way of matching would be the consider NNLLA and NNLO, but the NNLLA function $g_{3}$ is by now only known for $\tau$.
A new determination of $\alpha_{s}$ using ALEPH data is in progress. The analysis will follow the lines of the previous determination using pure NNLO predictions with a few improvements.

\section{Outlook}

The matching of NLLA and NNLO has improved the theoretical prediction of event shape distributions, but further improvement is possible by including the NNLL corrections into the calculations. These corrections are known only for $\tau$, where higher order logarithmic corrections have been computed~\cite{scetthrust} using soft-collinear effective theory (SCET). From these calculations one can extract the functions $g_{3}\left(\alpha_{s}L\right)$ and $g_{4}\left(\alpha_{s}L\right)$. The next step towards the further improvement in the extraction of $\alpha_{s}$ from event shape distributions could be to compute them for all six observables mentioned here. As shown in~\cite{scetthrust} the subleading logarithmic corrections can also account for about half of the discrepancy between parton-level theoretical predictions and hadron-level experimental data.

Improvements can also come from non-perturbative corrections. A very recent non-perturbative study for $\tau$ using a low-scale effective coupling~\cite{nonperturbativecorr} shows that non-perturbative $1/Q$ power corrections cause a shift in the distributions, which can account for an important part of the difference between parton-level distributions and hadron-level experimental data discussed in the previous section.

\section*{Acknowledgements}
We wish to thank the Swiss National Science
Foundation (SNF) which supported this work under contract 200020-117602.

% ****************************************************************************
% BIBLIOGRAPHY AREA
% ****************************************************************************

\begin{footnotesize}
% IF YOU DO NOT USE BIBTEX, USE THE FOLLOWING SAMPLE SCHEME FOR THE REFERENCES
% ----------------------------------------------------------------------------

% ----------------------------------------------------------------------------

% IF YOU USE BIBTEX,
% - DELETE THE TEXT BETWEEN THE TWO ABOVE DASHED LINES
% - UNCOMMENT THE NEXT TWO LINES AND REPLACE 'Name_Of_Your_BibFile'

%\bibliographystyle{unsrt}
%\bibliography{Name_Of_Your_BibFile}
% example of Name_Of_Your_BibFile.bib
% @Article{Turcato:2006ch,
%      author    = "Turcato, M.",
%  collaboration = "ZEUS and H1",
%      title     = "Lepton flavour violation and charmonium physics at HERA",
%      journal   = "Nucl. Phys. Proc. Suppl.",
%      volume    = "162",
%      year      = "2006",
%      pages     = "283-287",
%      SLACcitation  = "%%CITATION = NUPHZ,162,283;%%"
% }
%
% @Unpublished{Gogitidze:2007du,
%      author    = "Gogitidze, N.",
%  collaboration = "H1",
%      title     = "Prompt photons and particle momentum distributions at
%                   HERA",
%      year      = "2007",
%      note    = "hep-ex/0701033",
%      SLACcitation  = "%%CITATION = HEP-EX 0701033;%%"
% }

\end{footnotesize}

% ****************************************************************************
% END OF BIBLIOGRAPHY AREA
% ****************************************************************************

\end{document}